# Purcell-Enhanced Single-Photon Emission in the Telecom C-Band

*Jochen Kaupp,\* Yorick Reum, Felix Kohr, Johannes Michl, Quirin Buchinger, Adriana Wolf, Giora Peniakov, Tobias Huber-Loyola, Andreas Pfenning,\* and Sven Höfling*

**Purcell-enhanced quantum dot single-photon emission in the telecom C-band from InAs quantum dots inside circular Bragg grating cavities is shown. The InAs quantum dots are grown by means of molecular beam epitaxy on an InP substrate and are embedded into a quaternary $In_{0.53}Al_{0.23}Ga_{0.24}As$ membrane structure. In a post-growth flip-chip process with subsequent substrate removal and electron beam-lithography, circular Bragg grating ("bullseye") resonators are defined. Micro-photoluminescence studies of the devices at cryogenic temperatures of $T = 5$ K reveal individual quantum dot emission lines into a pronounced cavity mode. Time-correlated single-photon counting measurements under above-band gap excitation yield Purcell-enhanced excitonic decay times of $\tau = (180 \pm 3)$ ps corresponding to a Purcell factor of $F_P = (6.7 \pm 0.6)$. Pronounced photon antibunching with a background limited $g^{(2)}(0) = (0.057 \pm 0.004)$ is observed, which demonstrates that the light originated mostly from one single quantum dot.**

## 1. Introduction

As high-performant sources of single photons, epitaxial quantum dots (QDs) can be considered as a semiconductor launchpad for quantum photonic technologies,[1] with applications in quantum metrology, biology, and the foundations of quantum physics,[2] as well as in quantum communications and computing.[3] Of particular interest are quantum dot single-photon sources that emit light at wavelengths in the telecom C-band between 1530 – 1565 nm,[4] which would allow for seamless integration into fiber telecommunication networks at lowest losses.[5,6] Additionally

J. Kaupp, Y. Reum, F. Kohr, J. Michl, Q. Buchinger, A. Wolf, G. Peniakov,
T. Huber-Loyola, A. Pfenning, S. Höfling
Julius-Maximilians-Universität Würzburg
Physikalisches Institut, Lehrstuhl für Technische Physik
Am Hubland, 97074 Würzburg, Germany
E-mail: jochen.kaupp@uni-wuerzburg.de;
andreas.pfenning@physik.uni-wuerzburg.de

integrated silicon photonic components can be used at low loss with this wavelength.[7]

High single-photon emission probabilities[8] and indistinguishability[9–11] have been demonstrated with InAs QDs grown on GaAs. Unfortunately, this material platform is typically limited to wavelengths below 1300 nm,[12] with the exception of the exploitation of strain-relaxing layers like metamorphic buffers.[13–15] Metamorphic buffers as well as InP-based systems[16–19] have been used to reach wavelengths in the telecom C-band, and recent works on the GaSb/AlGaSb system[20] highlight GaSb QDs as another candidate for single-photon sources.

Due to the high refractive index contrast to air, QDs in bulk semiconductors with a refractive index > 3 offer poor extraction and collection efficiencies on the order of 1 %.[21] In $In_{0.53}Al_{0.23}Ga_{0.24}As$ for light collection optics with a moderate numerical aperture (NA) of NA = 0.4 the collection efficiency is about 1.6 %. Thus, QDs are typically implemented in resonator structures.[10,11,22–24] Micro-resonators such as photonic crystal cavities,[25] micropillar cavities,[26,27] or micro-disc resonators[28] offer high extraction efficiency combined with high emission rates due to Purcell enhancement. Within the past decade, circular Bragg grating, so-called "bullseye" resonators[29] have emerged as another promising micro resonator platform, because of a high Purcell enhancement over a broad wavelength range.[30–33] Single photons can be funneled into an almost Gaussian mode, with high extraction efficiencies over a broad spectral range.[32,34,35] QDs in bullseye resonators under resonant pumping have been demonstrated as near-ideal single-photon sources with high efficiencies, indistinguishability, and low multi-photon probability.[36,37] A variety of design studies demonstrate the aptitude of bullseye resonators as the ideal platform for III-V QD SPS at 1.55 μm.[32,35,38,39] Yet, comparatively little literature can be found on the experimental realization of bullseye SPS in the telecom C-band, with the exception of QDs on metamorphic buffer grown by metal organic vapor-phase epitaxy (MOVPE),[40,41] and MOVPE-grown InAs/InP QDs.[42]

Here, we study the implementation of $InAs/In_{0.53}Al_{0.23}Ga_{0.24}As$ QDs into bullseye resonators grown by molecular beam epitaxy (MBE). The optical and quantum optical properties of the QDs and bullseyes are studied. We show record values of the Purcell enhancement in the bullseye resonators as well as the low multi-photon contribution of the QD emission.

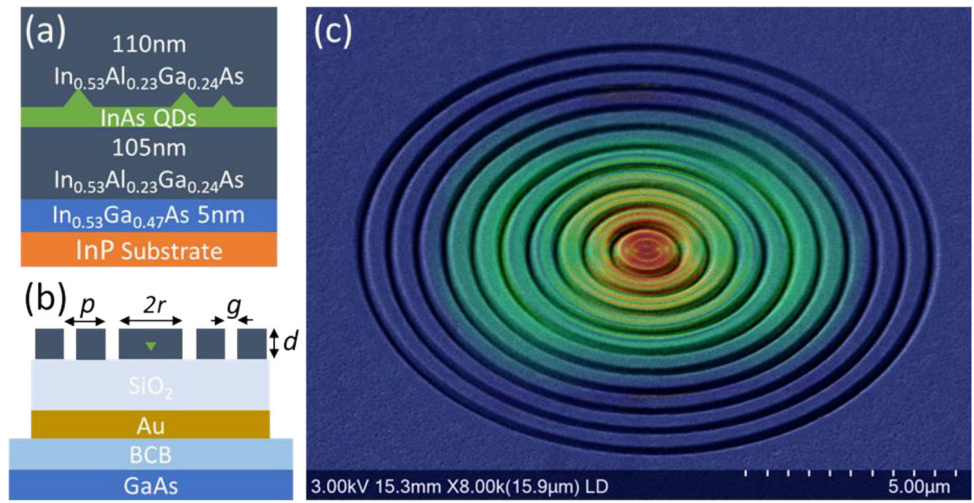

**Figure 1.** Sample design and fabrication. (a) Schematic layer structure after epitaxial growth. (b) Schematic layer structure after flip-chip and bullseye fabrication including the four bullseye design parameters. (c) Scanning electron microscopy (SEM) image of a fabricated bullseye resonator overlain with a logarithmic contour plot of the electric field intensity distribution of the fundamental resonator mode.

## 2. Sample Design and Fabrication

The QD heterostructure is depicted in **Figure 1** (a). The sample is grown by means of gas-source MBE on an n-type doped, (100)-oriented 2″-InP substrate. The group-III elements In, Ga, and Al are provided by thermal effusion cells, whereas the group-V elements As and P are provided by an $AsH_3/PH_3$ thermal cracker cell. Prior to the growth, a 5-minute oxide-desorption step is performed under constant phosphor flux. The growth process starts with a 200 nm $In_{0.52}Al_{0.48}As$ buffer that is lattice-matched to InP. Subsequently, 20 nm of an $In_{0.53}Ga_{0.47}As$ etch-stop layer is grown, followed by 90 nm of $In_{0.53}Al_{0.23}Ga_{0.24}As$ used to confine the InAs QDs. The quantum dot growth is performed under increased As-flux by depositing 2.5 monolayers of InAs with an Ostwald-ripening step during a 150 s growth interruption. The QDs are then covered by two monolayers of AlAs and encapsulated by 110 nm of $In_{0.53}Al_{0.23}Ga_{0.24}As$, which ends the growth sequence. A surface QD layer is grown on top of the final structure using the same growth conditions as for the buried QDs for morphological investigations of the QD shape and density. The morphology is investigated by means of atomic force microscopy. QD formation is observed with a spread in size and height. Emission in the telecom C-band is expected for QDs with a height exceeding 4.5 nm for which a density of $5 \times 10^8$ cm$^{-2}$ is found. Quantum dash formation is not observed.

Figure 1 (b) shows a cross-sectional sketch of the fabricated bullseye device. The nanofabrication starts with a plasma cleaning step, followed by sputter deposition of 540 nm of $SiO_2$. The metallic mirror is formed by depositing 100 nm of Au via electron beam evaporation. The sample is flipped and glued face-down onto a (10 × 10) mm$^2$ GaAs carrier substrate, spin-coated with an approximately 2 μm thick layer of the polymer Benzocyclobutene (BCB) serving as glue. The stacked sample is baked for 2.5 hours at a temperature of $T = 150°C$ under an applied pressure of ≈ 40 kPa. The InP substrate and the InAlAs buffer layer are removed by selective wet-etching in $HCl:H_3PO_4:CH_3COOH$ with a ratio of 1:1:2 (InP substrate), and $HCl:H_2O$ with a ratio of 3:1 (InAlAs buffer layer). High-resolution electron beam lithography with a PMMA resist is used to define bullseye resonators with varying parameters (Bragg grating period $p$, gap width $g$, and center disc radius $r$). The bullseye resonators are patterned in a grid with no deterministic relation between QD emission and the bullseye resonator. The pattern is transferred into the InAlGaAs membrane by reactive ion etching with an inductively coupled $Ar/Cl_2$ plasma. A scanning electron micrograph (SEM) of a nanofabricated QD bullseye resonator is shown in Figure 1 (c) for a device with $r = 767$ nm, $p = 660$ nm and $g = 140$ nm. The SEM is overlain with a finite difference time domain (FDTD) simulation of the electric field where the intensity of the field is plotted as a logarithmic contour plot, excited by a single dipole at the resonance frequency of the bullseye resonator and placed in its center.

Optimal device design parameters are identified by conducting 3D-FDTD numerical simulations. A constant low-temperature refractive index of $n = 3.3$ is assumed for the quaternary semiconductor. This value is based on a comparison of our experimental results with simulations and iterative adaption of the refractive index. Cross-sectional SEM confirms a membrane thickness of $d = (226 \pm 6)$ nm. Three co-dependent design parameters for which the resonator needs to be optimized remain $r$, $g$, and $p$. Exemplarily, **Figure 2** (a) shows the simulated Purcell factor on a colored contour plot as a function of the wavelength $\lambda$ and period $p$ (discrete period steps of 20 nm), for $r = 700$ nm, and $g = 140$ nm. The vertical dashed line indicates the target wavelength of 1550 nm. The maximal Purcell factor, full width at half maximum (FWHM), and center wavelength strongly depend on the period $p$, which demonstrates that the resonator dimensions need to be precisely matched with respect to each other. For $p = 640$ nm, a resonator mode at the target wavelength is found with a Purcell factor of $F_P = 27$. Deviating from this ideal value leads to a decreased Purcell enhancement, a resonance mode off-target, and increased FWHM.

To account for fabrication deviations and the unknown exact $In_{0.53}Al_{0.23}Ga_{0.24}As$ refractive index at cryogenic temperatures,

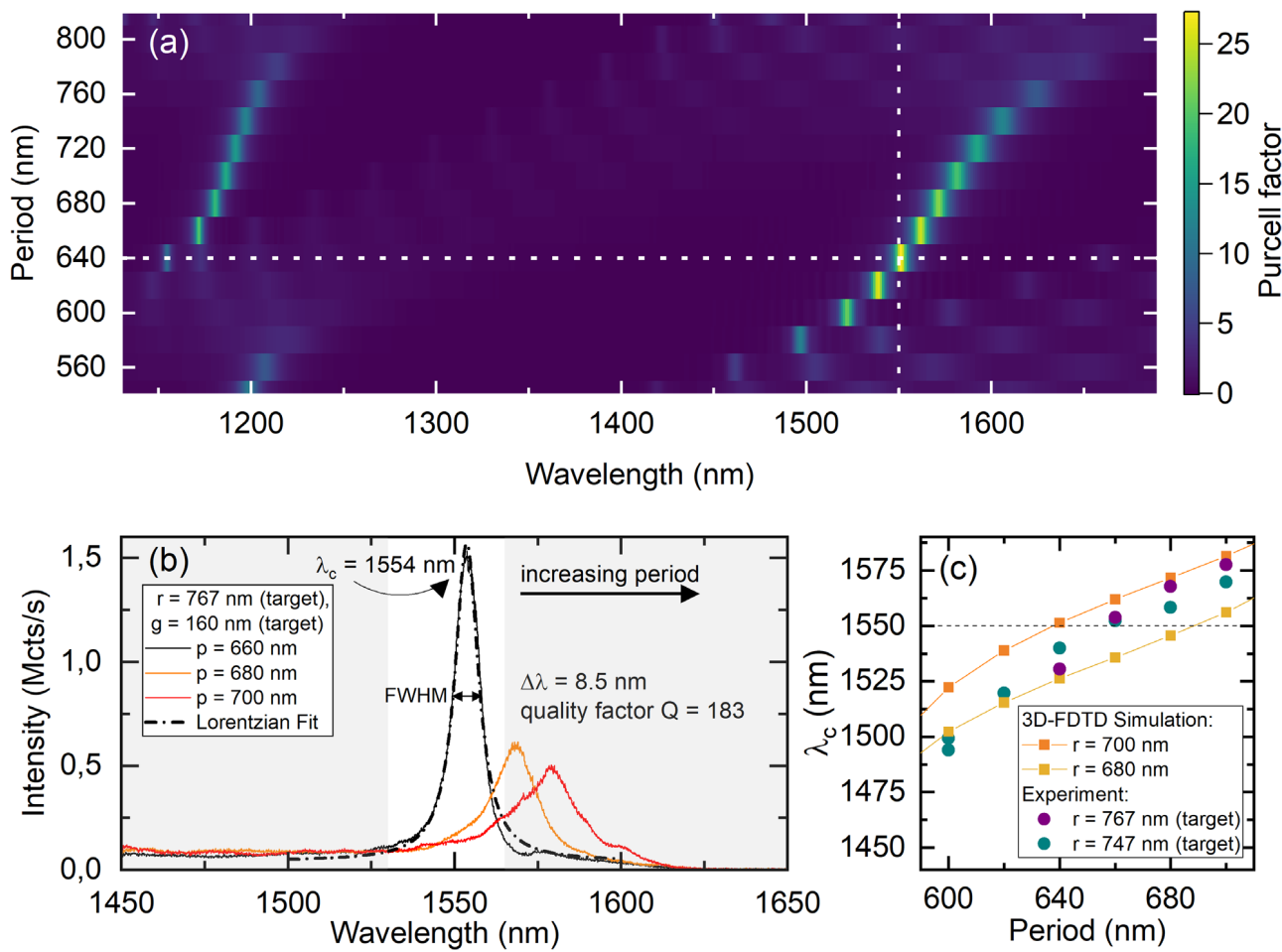

**Figure 2.** Simulated and experimental bullseye resonator modes. a) Colored contour plot of the simulated Purcell factor $F_P$ as a function of wavelength and Bragg grating period. The targeted wavelength of $\lambda = 1.55$ μm is indicated by the dotted white line. b) μPL spectrum for three different period values, showing the formation of a prominent resonator mode with a linewidth of 8.5 nm only for $p = 660$ nm. c) Comparison of simulations and experimental results. A systematic offset between experiment and simulation is observed and ascribed to a systematic over-etching.

the parameter space is studied experimentally by bracketing the three resonator parameters $p$, $r$, and $g$, and compared with the 3D-FDTD simulations. All values of $p$, $r$, and $g$ mentioned below are the nominal parameters. The optical properties of the fabricated devices are studied by means of micro-photoluminescence (μPL) spectroscopy at cryogenic temperatures of $T = (4.5 \pm 0.5)$ K in a helium flow cryostat. For excitation, a 660 nm continuous-wave laser is used. The μPL signal is collected with a NA = 0.4 guided to a 750 mm Czerny-Turner spectrometer and detected with a Nirvana 640 InGaAs near-infrared camera. At high excitation powers of $P = 70$ μW, all the QD lines are saturated and act as broadband gain medium for the resonator mode, which allows identification of the resonance wavelength and bandwidth (FWHM) of the bullseye resonator. Figure 2(b) shows the measured μPL spectra of three different bullseye resonators ($r = 767$ nm, and $g = 160$ nm) for increasing periods of $p = 660$, 680, and 700 nm as black, orange, and red line, respectively. For the device with $p = 660$ nm, a pronounced emission peak is found with a central wavelength $\lambda_c = 1554$ nm and a FWHM = 8.5 nm. This compares well with the simulated value of $FWHM_{sim} = 7.8$ nm. For increasing periods, the peak intensity decreases, while the cen-ter wavelength undergoes a redshift, and the FWHM increases, which is expected from the simulations. Figure 2 (c) shows a comparison of experimental results to the simulations. Experiments as well as simulations do show the same behaviour. The small deviation between theoretically predicted and experimentally obtained values could come from either the not precisely known refractive index of InAlGaAs at cryogenic temperatures and/or a structural deviation from the design. Indeed, SEM measurements of the etched structures show a systematic over-etching of the gap width of approximately 50 nm and a 50 nm reduced effective inner disc radius. Taking this into account, the experiment and simulations match well. A more in-depth study of the experimental result for bullseye parameters in the telecom C-band will be published in another work.

## 3. Purcell Enhancement and Quantum Optical Properties

The optical emission properties of the QD bullseye resonators are further studied by means of low-power μPL and spectrally resolved time-correlated single-photon counting. For excitation powers, $P < 5$ μW, the QD emission does not saturate and thus the resonator mode is not the only visible emission feature anymore. Individual QD emission lines can be observed on top of the still remaining resonator mode. An exemplary μPL spectrum taken at $P = 200$ nW is shown as black solid line in **Figure 3** (a).

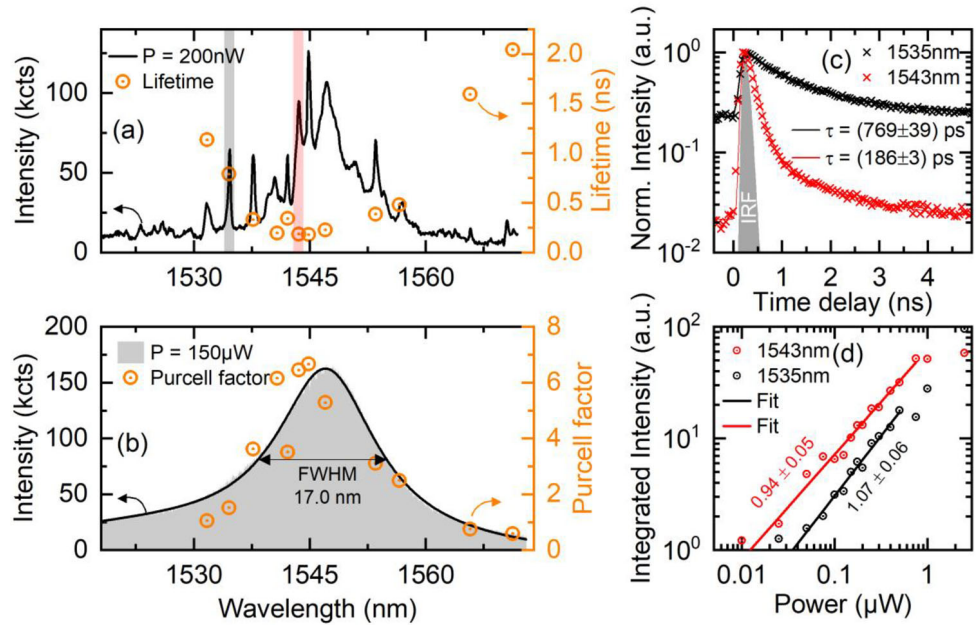

**Figure 3.** Lifetimes for different detuning to the cavity mode. a) Low-power μPL spectrum with various quantum dot emission lines. The lifetimes of the emission lines are shown as orange circles. b) High-power spectrum of the bullseye showing the resonator mode at $\lambda = 1547$ nm. The calculated Purcell factor with respect to bulk emission lifetime of $\tau = (1210 \pm 115)$ ps of quantum dot emission is plotted in orange at the respective wavelength of the emission. The highest value found here is $F_P = 6.7 \pm 0.6$. c) Lifetime measurements of the marked lines in (a). d) Power series of the marked lines.

For all following time-resolved single-photon counting, and second-order autocorrelation measurements the signal is filtered with a fiber-coupled band pass filter with a set FWHM of $(0.30 \pm 0.02)$ nm. The signal is then send to superconducting nanowire single-photon detectors (SNSPDs) via fiber. For these measurements, the QDs are excited with a pulsed laser with a wavelength of $\lambda = 720$ nm and an 80 MHz repetition rate.

To study the effect of the bullseye resonator mode on the lifetime, a device with multiple QD emission lines is located and studied, with design parameters $r = 740$ nm, $g = 140$ nm, and $p = 660$ nm. Due to the nondeterministic placement of the bullseye resonators with respect to the QD emission, not all resonators have single emission lines inside the resonator mode. This resonator does not have optimal parameters for the highest Purcell but is chosen due to the several well-separated single emission lines covering a wide spectral range of the bullseye resonance. The respective μPL spectrum for $P = 200$ nW is shown in Figure 3 (a). Due to the relatively high density of the QDs, various emission lines can be found while exciting within a single bullseye resonator. The lines are spectrally separated due to a size distribution of the QDs. The overall high density also leads to a large background emission into the bullseye resonator, even at low pump powers, when the QD emission does not saturate. Exciting with high excitation power $P = 150$ μW reveals the bullseye mode at $\lambda = 1547$ nm with a FWHM of 17.0 nm, shown in Figure 3 (b). Lifetimes for all resolvable QD emission lines are measured and plotted as orange circles in Figure 3 (a).

For the two red and grey marked emission lines in Figure 3 (a), the lifetime measurements are plotted on a logarithmic scale in Figure 3 (c). A bi-exponential decay can be observed in both lifetime measurements, which can be attributed to a different lifetime for the single-line emission and the large background emission underneath it. The background emission is expected to be spatially not aligned with the resonator mode, thus not Purcell enhanced. The reduction in the fast decay rate of the emission at $\lambda = 1543$ nm compared to emission at $\lambda = 1535$ nm is clearly visible. The slow decaying mechanism seems to be of a similar nature in both lines. The lifetimes below 300 ps are fitted with a numerical convolution of the measured instrument response function (IRF) and the expected bi-exponential decay. All other data in Figure 3 (a) is extracted from a bi-exponential decay function with an error function for the rise because the IRF is much shorter than the measured lifetimes.

To exclude overestimation of the Purcell factor by a factor of two by accidentally measuring biexciton recombination, a power series measurement is performed. The integrated intensity is plotted against the excitation power on log-log scale in Figure 3 (d), shown exemplarily for the two marked lines. Linear scaling is observed before the intensity saturates for all cases, which indicates excitonic emission from the respective QDs. Close to zero detuning at $\lambda = 1545$ nm, the lifetime is found to decrease down to $\tau = (180 \pm 3)$ ps compared to an off-resonant lifetime $\tau = (2.04 \pm 0.03)$ ns at $\lambda = 1571$ nm.

The decrease in lifetime originates from Purcell enhancement by the bullseye resonator. The emission lines are likely from separate QDs whose exact position with respect to the bullseye is unknown. The Purcell enhancement is strongly dependent on the exact position of the QD with respect to the optical resonator mode. Thus, it can not be concluded whether any of the QDs are experiencing optimal enhancement. Nevertheless, we observe a clear trend that correlates with the spectral detuning of the QDs to the resonator's resonance wavelength. Lifetimes of single-line emission in bulk are measured to get an estimate for lifetimes where the QD emission is not affected by the modification of the electric field. The commonly used approach to use QDs in the cavity that are spectrally detuned seems to overestimate the Purcell enhancement since the QDs electric field environment might deviate strongly from the electric field in bulk semiconductors. Averaging the lifetimes of several QD emission lines in the surrounding bulk material in a wavelength range from 1530 nm to 1565 nm results in an average lifetime of $\tau = (1210 \pm 115)$ ps, which is in very good agreement with the lifetime of $\approx 1280$ ps of MOVPE grown QDs on InP substrate[43] but lower than for previously studies of MOVPE grown QDs on InP of $\approx 2000$ ps.[42] Comparing this with the off-resonant lifetime $\tau = (2.04 \pm 0.03)$ ns at $\lambda = 1571$ nm in the bullseye resonator indicates that the emission from transitions spectrally detuned to the resonator wavelength could be suppressed.

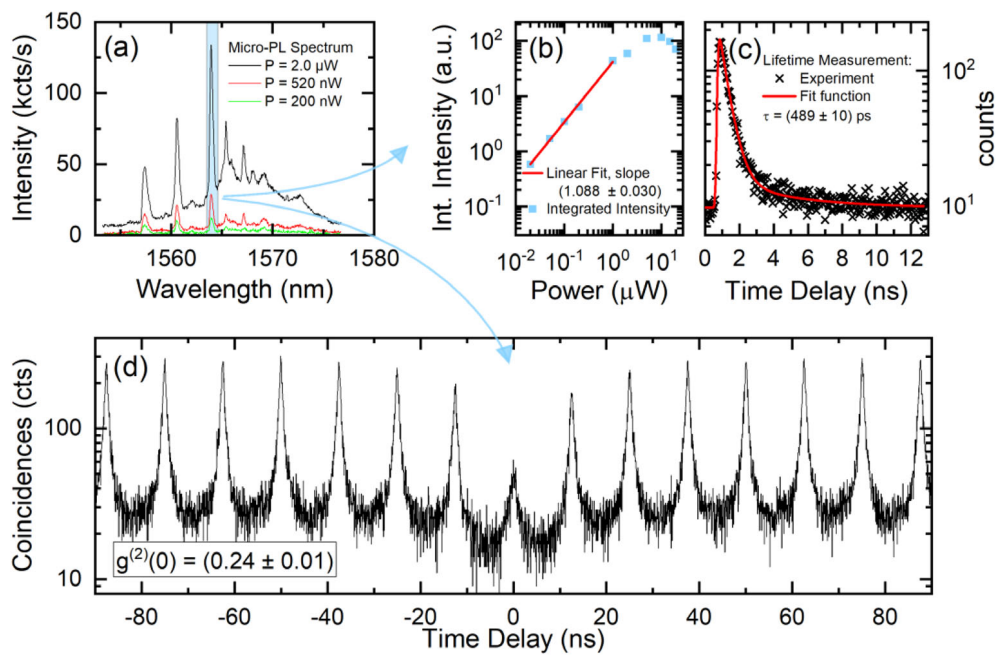

**Figure 4.** a) Low-power µPL spectrum of a bullseye resonator centered at $\lambda = 1567$ nm. The QD emission line at $\lambda = 1564$ nm is selected for further studies. b) Log-log plot of the intensity power series. The slope indicates an excitonic emission. c) Lifetime measurement with $\tau = (489 \pm 10)$ ps, which indicates a Purcell enhancement of about $F_P = 2.5 \pm 0.2$. d) Second-order autocorrelation measurement of the selected emission line on a logarithmic scale. At zero time delay antibunching is found, showing low multi-photon contribution from the studied source. The value of $g^{(2)}(0) = 0.24 \pm 0.1$ is in agreement with the expected value estimated from the signal-to-background ratio in the µPL spectra.

The calculated Purcell factor with respect to the bulk lifetime is plotted in Figure 3 (b) against the wavelength of the corresponding emission line. We calculate a maximum Purcell enhancement of $F_P = 6.7 \pm 0.6$ at $\lambda = 1545$ nm. The uncertainty of the Purcell factor originates from the spread of the reference bulk lifetimes.

The upper bound for the Purcell factor for this bullseye resonator can be estimated from comparison to the simulation. Simulated results for a resonator mode with a FWHM = 17.0 nm show a Purcell factor of $F_P = 10$. The QD emitting at $\lambda = 1545$ nm is likely spatially well aligned with the bullseye, because the Purcell factor is very close to the optimum, while the QD is spectrally still 2 nm detuned.

Although the above-studied bullseye resonator is very well suited to characterize the Purcell enhancement, it is not optimal for studying a single QD transition. To study single QD emission, we looked for areas on the sample with lower QD density. We exemplary study a device with design parameters $r = 760$ nm, $g = 140$ nm, and $p = 640$ nm. **Figure 4** (a) shows the µPL spectrum taken at different excitation powers. Three pronounced emission lines at $\lambda = 1564$ nm, $\lambda = 1561$ nm and $\lambda = 1557$ nm can be seen. The bullseye resonator mode is still apparent with a center wavelength $\lambda = 1567$ nm and FWHM of 10 nm, which adds to the background µPL signal. A power series of the emission line at $\lambda = 1564$ nm is taken and plotted in Figure 4 (b) on log-log scale. A power law coefficient of $1.088 \pm 0.030$ indicates excitonic (neutral or charged) nature of the QD charge complex. The decay time is measured via time-correlated single-photon counting and is shown in Figure 4 (c). A fit to a bi-exponential decay yields a decay time of $\tau = (489 \pm 10)$, which corresponds to a Purcell factor of $F_P = 2.5 \pm 0.2$.

The multi-photon emission probability is evaluated by a second-order autocorrelation measurement, which approximates the second-order autocorrelation function $g^{(2)}(\tau)$, where $\tau$ is the time delay between two detection events. The corresponding autocorrelation measurement is shown in Figure 4 (d) with a time resolution of 50 ps. Clear antibunching for zero-time delay indicates a low multi-photon contribution of the studied device with a dark count corrected $g^{(2)}(0) = 0.24 \pm 0.01$ (dark counts:

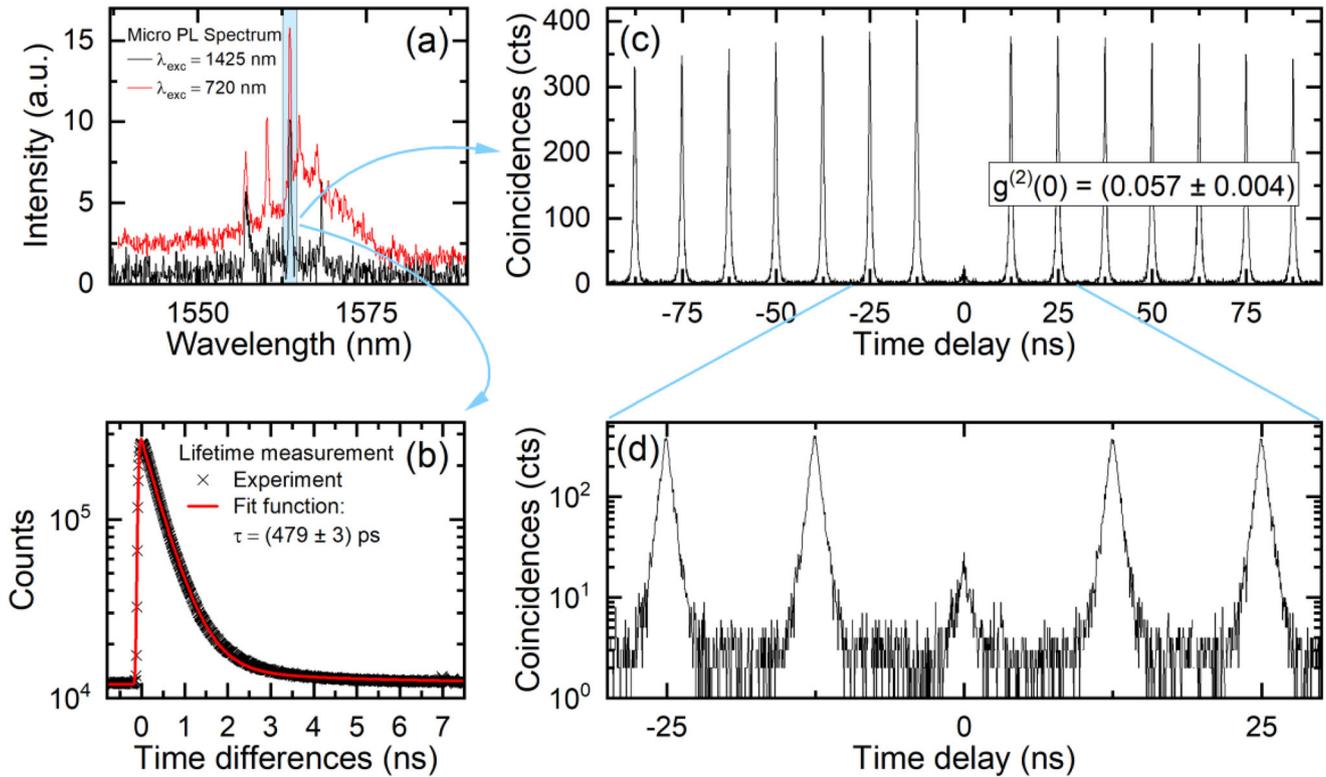

**Figure 5.** Quasi-resonant excitation via the p-shell a) µPL spectrum with p-shell and above band gap excitation. b) Lifetime measurement under p-shell excitation with $\tau = (479 \pm 3)$ ps. c) Second-order autocorrelation measurement of the selected emission line under p-shell excitation. Improved antibunching is found at zero-time delay compared to the above band gap excitation. d) Cut out of the second-order autocorrelation measurement plotted in logarithmic scale.

4.7 cts (50 ps)$^{-1}$; $g^{(2)}_{raw}(0) = 0.27 \pm 0.01$). A two-photon state yields $g^{(2)}(0) = 0.5$ and a single photon state $g^{(2)}(0) = 0$, thus a value of below 0.5 means that we measure a statistical mixture of single photons and more than one photon, where the single photon probability is dominating. The $g^{(2)}(0)$ is background emission limited as is apparent from the measured µPL spectrum (see Figure 4 (a)). Indeed, comparing for $P = 200$ nW the line intensity versus background intensity from the bullseye mode, we would expect a minimal achievable value for this line and device of $g^{(2)}(0) = 0.22$. The coincidences near zero-time delay are decreased. This can be ascribed to long timescale single charge filling mechanisms and/or dark state spin flips.[44]

The same device is studied under quasi-resonant excitation inside a closed cycle cryostat at $T = 2$ K. The signal is fiber-coupled and measured with the same detectors as the measurements in Figure 4. A wavelength scan of the excitation laser reveals p-shell excitation at $\lambda = 1425$ nm. The µPL spectrum under p-shell excitation is shown in **Figure 5** (a) as a solid black line and compared with the spectrum for above band gap excitation (solid red line). Under p-shell excitation, the background emission intensity into the bullseye optical mode is strongly reduced and only discrete QD emission lines remain. Time-correlated single-photon counting is performed to study the excitonic decay time (Figure 5 (b)) and probe the multi-photon emission probability (Figure 5 (c,d)). A bi-exponential fit yields a decay time of $\tau = (479 \pm 3)$ ps which corresponds to $F_P = 2.5 \pm 0.2$ and is comparable to the Purcell factor for above band gap excitation.

The second-order autocorrelation measurement is performed with 50 ps time resolution and is shown in Figure 5 (c). Improved antibunching is found at zero-time delay with a dark count corrected $g^{(2)}(0) = 0.057 \pm 0.004$ (dark counts: 3.2 cts (50 ps)$^{-1}$; $g^{(2)}_{raw}(0) = 0.077 \pm 0.004$). Figure 5 (d) shows a cut out of the second-order autocorrelation measurement in logarithmic scale.

## 4. Conclusion

InAs/InAlGaAs semiconductor QDs emitting in the telecom-C band were grown by MBE and integrated into bullseye resonators. The experimental data of the bullseye resonators can be reproduced by the simulation. We present record values of the Purcell factor of $F_P = 6.7 \pm 0.6$ for QDs at this wavelength range in any material system and cavity design. We expect the real Purcell factor to be even higher. The interfaces of the membrane in proximity to the quantum dots are expected to already have an impact on the lifetime,[45] resulting in lower lifetimes and underestimating our Purcell factor. This is still the best approximation for the unmodified lifetime. A $g^{(2)}(0) = 0.057 \pm 0.004$ demonstrates that we can study the emission from predominantly single QDs embedded in the bullseye resonators. This highlights our system as an excellent candidate for single-photon sources in the telecom C-band. Due to the nondeterministic positioning of the quantum dots as well as their high spatial density, we expect to improve these results in the future with deterministic placement[31] into further optimized structures on lower QD density samples.

## Acknowledgements

The authors acknowledge the support of the state of Bavaria and the German Ministry for Research and Education (BMBF) within Project PhotonQ (FKZ: 13N15759). Tobias Huber-Loyola acknowledges funding from the BMBF through the Quantum Futur (FKZ: 13N16272) initiative. The authors acknowledge expert technical assistance from Margit Wagenbrenner, Monika Emmerling, and Sabrina Estevam, and also thank Tobias Heindel from TU Berlin for helpful and insightful discussions.

Open access funding enabled and organized by Projekt DEAL.

## Conflict of Interest

The authors declare no conflict of interest.

## Author Contributions

Jochen Kaupp: Formal analysis; Investigation; Resources; Visualization; Writing – original draft; Writing – review & editing. Yorick Reum: Formal analysis; Investigation; Resources; Visualization; Writing – original draft; Writing – review & editing. Felix Kohr: Formal analysis; Investigation; Resources; Visualization; Writing – original draft; Writing – review & editing. Johannes Michl: Investigation; Writing – review & editing. Quirin Buchinger: Resources; Writing – review & editing. Adriana Wolf: Resources. Giora Peniakov: Investigation; Writing – review & editing. Tobias Huber-Loyola: Conceptualization; Formal analysis; Funding acquisition; Methodology; Project administration; Supervision, Validation; Writing – original draft; Writing – review & editing. Andreas Pfenning: Conceptualization; Formal analysis; Investigation; Funding acquisition; Methodology; Project administration; Supervision, Validation; Visualization; Writing – original draft; Writing – review & editing. Sven Höfling: Conceptualization; Funding acquisition; Methodology; Project administration; Supervision, Validation; Writing – review & editing.

## Data Availability Statement

The data that support the findings of this study are available from the corresponding author upon reasonable request.

## Keywords

bullseye resonators, III-V semiconductors, quantum dots, single-photon sources, telecom C-band

[1] X. Zhou, L. Zhai, J. Liu, *Photonics Insights* **2022**, *1*, R07.
[2] C. Couteau, S. Barz, T. Durt, T. Gerrits, J. Huwer, R. Prevedel, J. Rarity, A. Shields, G. Weihs, *Nat Rev Phys* **2023**, *5*, 354.
[3] C. Couteau, S. Barz, T. Durt, T. Gerrits, J. Huwer, R. Prevedel, J. Rarity, A. Shields, G. Weihs, *Nat Rev Phys* **2023**, *5*, 326.
[4] T. Miyazawa, K. Takemoto, Y. Sakuma, S. Hirose, T. Usuki, N. Yokoyama, M. Takatsu, Y. Arakawa, *Jpn. J. Appl. Phys.* **2005**, *44*, L620.
[5] N. Gisin, R. Thew, *Nat. Photonics* **2007**, *1*, 165.
[6] C. Schimpf, M. Reindl, F. Basso Basset, K. D. Jöns, R. Trotta, A. Rastelli, *Appl. Phys. Lett.* **2021**, *118*, 100502.
[7] S. Y. Siew, B. Li, F. Gao, H. Y. Zheng, W. Zhang, P. Guo, S. W. Xie, A. Song, B. Dong, L. W. Luo, C. Li, X. Luo, G.-Q. Lo, *J. Lightwave Technol.* **2021**, *39*, 4374.
[8] J. Neuwirth, F. Basso Basset, M. B. Rota, J.-G. Hartel, M. Sartison, S. F. Covre Da Silva, K. D. Jöns, A. Rastelli, R. Trotta, *Phys. Rev. B* **2022**, *106*, L241402.
[9] Y.u-M. He, Y.u He, Y.u-J. Wei, D. Wu, M. Atatüre, C. Schneider, S. Höfling, M. Kamp, C.-Y. Lu, J.-W. Pan, *Nature Nanotech* **2013**, *8*, 213.
[10] N. Somaschi, V. Giesz, L. De Santis, J. C. Loredo, M. P. Almeida, G. Hornecker, S. L. Portalupi, T. Grange, C. Antón, J. Demory, C. Gómez, I. Sagnes, N. D. Lanzillotti-Kimura, A. Lemaître, A. Auffeves, A. G. White, L. Lanco, P. Senellart, *Nat. Photonics* **2016**, *10*, 340.
[11] X. Ding, Y.u He, Z.-C. Duan, N. Gregersen, M.-C. Chen, S. Unsleber, S. Maier, C. Schneider, M. Kamp, S. Höfling, C.-Y. Lu, J.-W. Pan, *Phys. Rev. Lett.* **2016**, *116*, 20401.
[12] M. Paul, J. Kettler, K. Zeuner, C. Clausen, M. Jetter, P. Michler, *Appl. Phys. Lett.* **2015**, *106*, 122105.
[13] E. S. Semenova, R. Hostein, G. Patriarche, O. Mauguin, L. Largeau, I. Robert-Philip, A. Beveratos, A. Lemaître, *J. Appl. Phys.* **2008**, *103*, 103533.
[14] S. L. Portalupi, M. Jetter, P. Michler, *Semicond. Sci. Technol.* **2019**, *34*, 53001.
[15] P. A. Wronski, P. Wyborski, A. Musial, P. Podemski, G. Sek, S. Höfling, F. Jabeen, *Materials* **2021**, *14*, 5221.
[16] D. Fuster, A. Rivera, B. Alén, P. Alonso-González, Y. González, L. González, *Appl. Phys. Lett.* **2009**, *94*, 133106.
[17] M. Benyoucef, M. Yacob, J. P. Reithmaier, J. Kettler, P. Michler, *Appl. Phys. Lett.* **2013**, *103*, 162101.
[18] M. Yacob, J. P. Reithmaier, M. Benyoucef, *Appl. Phys. Lett.* **2014**, *104*, 022113.
[19] T. Müller, J. Skiba-Szymanska, A. B. Krysa, J. Huwer, M. Felle, M. Anderson, R. M. Stevenson, J. Heffernan, D. A. Ritchie, A. J. Shields, *Nat. Commun.* **2018**, *9*, 862.
[20] J. Michl, G. Peniakov, A. Pfenning, J. Hilska, A. Chellu, A. Bader, M. Guina, S. Höfling, T. Hakkarainen, T. Huber-Loyola, Strain-free GaSb quantum dots as single-photon sources in the telecom S-band, **2023**, arXiv:2305.04384v1.
[21] W. L. Barnes, G. Björk, J. M. Gérard, P. Jonsson, J. A. E. Wasey, P. T. Worthing, V. Zwiller, *Eur. Phys. J. D* **2002**, *18*, 197.
[22] M. Pelton, C. Santori, J. Vuc?Kovic, B. Zhang, G. S. Solomon, J. Plant, Y. Yamamoto, *Phys. Rev. Lett.* **2002**, *89*, 233602.
[23] A. Dousse, J. Suffczynski, A. Beveratos, O. Krebs, A. Lemaître, I. Sagnes, J. Bloch, P. Voisin, P. Senellart, *Nature* **2010**, *466*, 217.
[24] L. Ginés, M. Moczala-Dusanowska, D. Dlaka, R. HosaK, J. R. Gonzales-Ureta, J. Lee, M. Jezek, E. Harbord, R. Oulton, S. Höfling, A. B. Young, C. Schneider, A. Predojevic, *Phys. Rev. Lett.* **2022**, *129*, 33601.
[25] Z. Lin, J. Vuckovic, *Phys. Rev. B* **2010**, *81*, 35301.
[26] O. Gazzano, S. Michaelis De Vasconcellos, C. Arnold, A. Nowak, E. Galopin, I. Sagnes, L. Lanco, A. Lemaître, P. Senellart, *Nat. Commun.* **2013**, *4*, 1425.
[27] S. Unsleber, C. Schneider, S. Maier, Y.u-M. He, S. Gerhardt, C.-Y. Lu, J.-W. Pan, M. Kamp, S. Höfling, *Opt. Express* **2015**, *23*, 32977.
[28] J. Yang, S. Shi, X. Xie, S. Wu, S. Xiao, F. Song, J. Dang, S. Sun, L. Yang, Y. Wang, Z.i-Y. Ge, B.-B. Li, Z. Zuo, K. Jin, X. Xu, *Opt. Express, OE* **2021**, *29*, 14231.
[29] J. Scheuer, A. Yariv, *IEEE J. Quantum Electron.* **2003**, *39*, 1555.
[30] M. Davanço, M. T. Rakher, D. Schuh, A. Badolato, K. Srinivasan, *Appl. Phys. Lett.* **2011**, *99*, 041102.
[31] L. Sapienza, M. Davanço, A. Badolato, K. Srinivasan, *Nat. Commun.* **2015**, *6*, 7833.
[32] A. Barbiero, J. Huwer, J. Skiba-Szymanska, T. Müller, R. M. Stevenson, A. J. Shields, *Opt. Express* **2022**, *30*, 10919.
[33] Q. Buchinger, S. Betzold, S. Höfling, T. Huber-Loyola, *Appl. Phys. Lett.* **2023**, *122*, 111110.
[34] W. B. Jeon, J. S. Moon, K.-Y. Kim, Y.-H.o Ko, C. J. K. Richardson, E. Waks, J.-H. Kim, *Adv. Quant. Technol.* **2022**, *5*, 2200022.

[35] L. Rickert, F. Betz, M. Plock, S. Burger, T. Heindel, *Opt. Express* **2023**, *31*, 14750.

[36] H. Wang, Y.u-M. He, T.-H. Chung, H. Hu, Y. Yu, S.i Chen, X. Ding, M.-C. Chen, J. Qin, X. Yang, R.-Z.e Liu, Z.-C. Duan, J.-P. Li, S. Gerhardt, K. Winkler, J. Jurkat, L.-J. Wang, N. Gregersen, Y.-H. Huo, Q. Dai, S. Yu, S. Höfling, C.-Y. Lu, J.-W. Pan, *Nat. Photonics* **2019**, *13*, 770.

[37] J. Liu, R. Su, Y. Wei, B. Yao, S. F. C. D.a Silva, Y. Yu, J. Iles-Smith, K. Srinivasan, A. Rastelli, J. Li, X. Wang, *Nat. Nanotechnol.* **2019**, *14*, 586.

[38] L. Bremer, C. Jimenez, S. Thiele, K. Weber, T. Huber, S. Rodt, A. Herkommer, S. Burger, S. Höfling, H. Giessen, S. Reitzenstein, *Opt. Express* **2022**, *30*, 15913.

[39] D. Bauch, D. Siebert, K. D. Jöns, J. Förstner, S. Schumacher, On-demand indistinguishable and entangled photons using tailored cavity designs, **2023**, arXiv:2303.13871v2.

[40] R. Sittig, C. Nawrath, S. Kolatschek, S. Bauer, R. Schaber, J. Huang, P. Vijayan, P. Pruy, S. L. Portalupi, M. Jetter, P. Michler, *Nanophotonics* **2022**, *11*, 1109.

[41] C. Nawrath, R. Joos, S. Kolatschek, S. Bauer, P. Pruy, F. Hornung, J. Fischer, J. Huang, P. Vijayan, R. Sittig, M. Jetter, S. L. Portalupi, P. Michler, *Adv Quantum Technol.* **2023**, *6*, 2300111.

[42] P. Holewa, E. Zięba-Ostój, D. A. Vajner, M. Wasiluk, B. Gaál, A. Sakanas, M. Burakowski, P. Mrowiński, B. Krajnik, M. Xiong, A. Huck, K. Yvind, N. Gregersen, A. Musiał, T. Heindel, M. Syperek, E. Semenova, Scalable quantum photonic devices emitting indistinguishable photons in the telecom C-band, **2023**, arXiv:2304.02515v1.

[43] D. A. Vajner, P. Holewa, E. Zięba-Ostój, M. Wasiluk, M. v. Helversen, A. Sakanas, A. Huck, K. Yvind, N. Gregersen, A. Musiał, M. Syperek, E. Semenova, T. Heindel, On-demand Generation of Indistinguishable Photons in the Telecom C-Band using Quantum Dot Devices, **2023**, arXiv:2306.08668v1.

[44] C. Santori, D. Fattal, J. Vuckovic, G. S. Solomon, E. Waks, Y. Yamamoto, *Phys. Rev. B* **2004**, *69*, 205324.

[45] S. Stobbe, J. Johansen, P. T. Kristensen, J. M. Hvam, P. Lodahl, *Phys. Rev. B* **2009**, *80*, 155307.